\documentclass[aps,prc,amsmath,superscriptaddress,twocolumn,showpacs,captions=nooneline]{revtex4-1} 
\usepackage{graphicx,amsmath,amssymb,bm,color,inputenc,fontenc,mathrsfs}
\usepackage{epstopdf}
\usepackage{float}
\usepackage[caption = false]{subfig}
\usepackage{bm}

\begin{document}

\title[title]{Symmetry restoration in  mixed-spin paired heavy nuclei}
\author{Ermal Rrapaj}
\affiliation{Department of Physics, University of Guelph, Guelph, Ontario N1G 2W1, Canada}
\author{A. O. Macchiavelli}
\affiliation{Nuclear Science Division, Lawrence Berkeley National Laboratory, Berkeley, CA 94720, USA}
\author{Alexandros Gezerlis}
\affiliation{Department of Physics, University of Guelph, Guelph, Ontario N1G 2W1, Canada}

\begin{abstract}
The nature of the nuclear pairing condensates in heavy nuclei, specifically neutron-proton (spin-triplet), versus
identical-particle (spin-singlet) pairing has been an active area of research for quite some time.
In this work, we  probe three candidates that should display spin-triplet, spin-singlet, and
mixed-spin pairing. Using theoretical approaches such as the gradient method and symmetry restoration techniques, 
we find the ground state of these nuclei in Hartree-Fock-Bogoliubov theory and compute ground state to ground state pair-transfer
amplitudes to neighboring isotopes while simultaneously projecting to specific particle number and nuclear spin values.
We identify specific reactions for future experimental research that could shed light on spin-triplet and mixed-spin pairing.

\end{abstract}
\maketitle
\section{Introduction}
\label{section:intro}
The presence of pairing in atomic nuclei has been established for more than five decades \cite{Bohr1958}. Extensive experimental data on nuclear properties:
even-even excitation gaps, binding-energy differences, moments of inertia, onset of deformation, two-nucleon transfer reactions, etc. can be explained by the presence of
neutron-neutron (nn) and proton-proton (pp) Bardeen-Cooper-Schrieffer-like (BCS-like) pairing \cite{Dean2003, Brink2010, Broglia2013}.

For most known nuclei, with neutron excess, the ground state consists of $nn$ and $pp$ ($j=0$, $t=1$ ) pairs coupled to angular momentum $J=0$.
For nuclei with comparable number of neutrons and protons, the nucleons near the Fermi surface should occupy
identical orbitals and $np$ pairing should be present. Due to the Pauli
exclusion principle, isospin-singlet and isoscalar ($t=0$) is associated with spin-triplet ($s=1$) pairing, and vice versa.

The elusive spin-triplet pairing in nuclei has been both an experimental and theoretical puzzle over the decades \cite{Frauendorf2014}. 
Charge independence of the nuclear force should lead to both ($j=0,t=1$) $nn$ and $pp$ pairing on equal footing with ($j=0,t=1$) $np$ pairing
for nuclei with $N\approx Z$.
In addition, the existence of the deuteron as a $J^{\pi}=1^+$ bound state and low-energy scattering data \cite{Stoks1993} indicate that
the strength of the interaction is stronger in the isoscalar channel in comparison with nucleons coupled to isospin 1.
The natural conclusion from this observation is the expectation to find isospin-singlet, spin-triplet pairing in nuclei,
in the form of a {\sl quasi}deuteron condensate. 

Neutron-proton pair correlations have been studied by analyzing the results of large-scale shell-model calculations 
\cite{Engel1996, Engel1998, Langanke1995, Langanke1996,Langanke1996_2, Langanke1997, Dean1997, Poves1998, MartinezPinedo1998}.
The spin-orbit interaction tends to suppress spin-triplet pairing \cite{Poves1998, Baroni2010}, and nuclear deformation also plays a competitive role
and therefore needs to be treated in detail~\cite{Bonatsos2017}. In the case of $N \approx\ Z$, and large atomic number, if one assumes spherical symmetry, it is reasonable to expect
this type of pairing.

However, in finite systems, pairing can be difficult to 
define, and many proxies have been used in the literature ~\cite{Engel1996, Engel1998, Langanke1995, Langanke1996_2, Langanke1997}. 
The energy competition between the spin-singlet and spin-triplet states has also been studied~\cite{Tanimura2014}.
The most direct measure would be to calculate the pair-transfer reaction probabilities~\cite{Brink2010,Broglia2013} and here we calculate the pair-transfer amplitude
in the framework of Hartree-Fock-Bogoliubov (HFB) theory.

The Hartree-Fock-Bogoliubov approach is a versatile tool that can describe a large number of many-nucleon problems where pairing is important \cite{Ring1980}.
The basics of the HFB formalism are covered in Sec. \ref{section:HGS}.
Pairing studies in nuclear physics have included an isovector pairing field, an isoscalar pairing field, and coexisting ($t = 0,1$) pairing fields for $N = Z$, as well as general nucleon numbers  \cite{Camiz1965,Camiz1966_2, Ginocchio1968,Goswami1964, Goswami1965, Chen1967, Goodman1968, Wolter1970, Wolter1971}. 
More recently, a mixed-spin pairing ground state was found to be energetically favorable, in the context of HFB theory, for the case
of heavy nuclei~\cite{Gezerlis2011,Bulthuis2016} (see also Ref.~\cite{Bertsch2010}).

In this work, we focus our attention on the $A\geq 130$ region close to the proton dripline.
In Ref.~\cite{Gezerlis2011} many candidates where $t=0$ pairing could be present were found in this area.
While we are aware that transfer reaction studies on these nuclei are currently not possible, this part of the nuclear chart could
be accessible to experimental research via selective studies of fusion-evaporation reactions. Thus our findings, based on the analysis of two-nucleon overlaps,
can guide the experimental program to those nuclei where the presence of a spin-triplet pairing phase near the ground state is more probable.

The first step, then, is finding the ground state for a given nucleus. In practice, particle number
and nuclear spin  are not conserved and need to be restored. 
Employing the gradient method developed in Ref.~\cite{Robledo2011} we find the minimal-energy wave function.   
This method allows one to constrain the expectation value of particle number and the amplitudes of various pairing channels.
We do so to explore how various constraints impact not only the energy of the ground state but also its composition in terms of
eigenstates of the symmetry operators under consideration.

Symmetry restoration can be a nontrivial task.
In the past, various formulas based on determinants have been used, which suffer from a sign ambiguity \cite{Onishi1966};
and various approximations to overcome it have been employed  \cite{Neergard1983, Haider1992, Donau1998, Robledo2009}.
Ambiguity-free formulations have been recently developed \cite{Bertsch2012, Avez2012, Oi2012}. 
We make use of the expressions derived in Ref.~\cite{Bertsch2012}, which do not have the shortcoming mentioned.

As found in Refs.~\cite{Gezerlis2011,Bulthuis2016}, there are nuclei where one type of pairing dominates, like spin-triplet in ${}^{132}_{66}$Dy, or spin-singlet in  
${}^{132}_{60}$Nd. Also nuclei with coexistence of both types are present in the nuclear chart, like the so-called mixed-spin
pairing in ${}^{132}_{64}$Gd.
The distributions of the states of good quantum numbers for the ground state of each of these three nuclei are 
analyzed in Secs. \ref{subsection:part_num}, \ref{subsection:nuc_spin}, and \ref{subsection:nuc_spin_n}.

Another area of  investigation is how pair-transfer cross sections (probabilities) compare 
in ground state to ground state transitions \cite{Grasso2012}, an observable that could be considered as the {\sl smoking gun} to disentangle the two effects.
We compute various transitions from the neighboring isotopes of the three nuclei mentioned, while simultaneously carrying out a symmetry projection.

In this paper, our goal is twofold: (i) To confirm the nature of the ground state condensates survives after projection and, (ii) For future studies,  
 to find the most promising pair-transfer reactions for each case. A detailed discussion can be found in Sec.
\ref{section:emission}, and we draw our conclusions in the last section.

\section{The HFB Formalism}
\label{section:HGS}
The HFB theory is based on a variational principle for the energy of the ground state of the system.
The many-body wave function is varied in the space of Slater determinants of quasiparticles defined by the Bogoliubov transformation.
The ``effective'' Hamiltonian in this theory consists of one-body and two-body operators, which we write in second quantization language, 
in terms of spin-half particle operators, as
\begin{equation}
 \begin{split}
  \hat{H}=\sum_{i,j}t_{ij}c^{\dagger}_ic_j+\frac{1}{4}\sum_{i,j,k,l}\overline{v}_{ijkl}c^{\dagger}_ic^{\dagger}_jc_lc_k
 \end{split}
\end{equation}
The one body potential used in this work is of Wood-Saxon shape including contributions from spin-orbit interactions,
\begin{equation}
 \begin{split}
  v(\boldsymbol{r})=&V_{WS}f({r})-(\bm{L}\cdot \bm{S})\frac{V_{SO}}{r} \frac{df}{dr}\\
  f(r)=&[1+e^{(r-R)/a}]^{-1}
 \end{split}
\end{equation}
and the two body interaction is a contact term for each of the pairing channels given in Table \ref{tab:pairing},
\begin{equation}
 \begin{split}
  V(\bm{r}_1,\bm{r}_2)=&\sum_{\alpha}^{6}v_{\alpha}P_{L=0}P_{\alpha}\delta^{3}(\bm{r}_1-\bm{r}_2)\\
   =&\frac{1}{4}\bigg(3v_t + v_s + (v_t - v_s) \bm{\sigma}_1 \cdot \bm{\sigma}_2\bigg)\\
   &\times \delta^{3}(\bm{r}_1-\bm{r}_2)P_{L=0}
 \end{split}
\end{equation}
The numerical values for the parameters $v_s$  and $v_t$ are 300 and 450 MeV respectively, taken from Ref.~\cite{Bulthuis2016}.  
The Bogoliubov transformation from particle to quasiparticle space is defined as follows:
\begin{equation}
 \begin{split}
  \begin{pmatrix} { {\bm \beta}} \\ {\boldsymbol \beta}^{\dagger} \end{pmatrix} = \begin{pmatrix} U^{\dagger} && V^{\dagger} \\
                                                                           V^T && U^T
                                                           \end{pmatrix}
  \begin{pmatrix} \bm{c} \\ \bm{c}^{\dagger}\end{pmatrix}
 \end{split}
\end{equation}
As a result, the Hamiltonian can be expressed in the new basis,
\begin{equation}
 \begin{split}
  \hat{H}=H^{00}+\bm \beta^{\dagger}H^{11}\bm\beta + \frac{1}{2} \bm \beta^{\dagger} H^{20} \bm \beta^{\dagger}+ \ldots
 \label{eq:H}
 \end{split}
\end{equation}
where the superscripts count the number of creation and annihilation operators of quasiparticles. A more detailed explanation of the various terms
appearing in Eq.~(\ref{eq:H}) can be found in Ref.~\cite{Bulthuis2016}.

\subsection{General features of the ground state}
The ground state wave function used in this work is defined as follows:
\begin{equation}
 \begin{split}
  |\Phi\rangle =& \text{pf}(U^{\dagger}V^{*})\ \text{exp}\bigg[\frac{1}{2} {(VU^{-1})^{*}}_{i j} c^{\dagger}_i c^{\dagger}_j\bigg]|0\rangle\\
  \label{eq:wvfn}
 \end{split}
\end{equation}
where $\text{pf}()$ is the Pfaffian of the matrix, and $|0\rangle$ is the reference vacuum state. The three main isotopes investigated here share the same reference
vacuum state, and the same quasiparticle basis, which technically is infinite. Different isotopes occupy different subspaces, and when their overlap is calculated,
an augmented subspace which encompasses both nuclei is used \cite{Robledo2011_2}.
The minimization of the energy is performed through the gradient method described in Ref.~\cite{Robledo2011} subject to neutron and proton number constraints.
In addition, the various nucleon pairing channels can be constrained \cite{Bulthuis2016}, and the constrained Hamiltonian is
\begin{equation}
 \begin{split}
  \hat{H}_c=\hat{H}-\sum_{\alpha}\lambda_{\alpha} \hat{Q}_{\alpha}
 \end{split}
\end{equation}
The parameters $\lambda_{\alpha}$ are analogous to Lagrange multipliers and the operators $Q_{\alpha}$ are particle number, pairing amplitudes, etc.
In this sense, this formulation employs the grand canonical ensemble.

As already mentioned in the Introduction, the three representative isotopes analyzed here are ${}^{132}_{60}$Nd, ${}^{132}_{64}$Gd ,
and ${}^{132}_{66}$Dy, taken from Ref.~\cite{Bulthuis2016}.
While we find a distribution of eigenstates
with specific quantum numbers in the ground state, we enforce this distribution to be highly peaked at the target isotope.

\begin{table}[]
\centering
\begin{tabular}{c||cccccc}
\hline
& 1 & 2 & 3 & 4 & 5 & 6 \\ \hline \hline
$(S,\ S_z)$ & (0, 0) & (0, 0) & (0, 0) & (1, 1) & (1, 0) & (1,-1) \\\hline
$ (T,\ T_z) $& (1, 1) & (1, 0) & (1, -1) & (0, 0) & (0, 0) &  (0, 0) \\\hline
\end{tabular}
\caption{The 6 spin-isospin pairing channels}
\label{tab:pairing}
\end{table}
All the various possible pairing channels are given in Table \ref{tab:pairing}.
In Table \ref{table:Ecorr} we report the correlation energy, 
the energy difference between the unpaired ground state, and the one without any suppression of pairing,
found for each isotope subject to pairing constraints.
Since the present calculations are at the mean-field level, the results are to be understood
more as a qualitative rather than quantitative representation of the ``physical'' ground state.  
\begin{table}[]
\centering
\begin{tabular}{c||ccc}
\hline
&  ${}^{132}_{66}$Dy & ${}^{132}_{64}$Gd & ${}^{132}_{60}$Nd  \\ \hline \hline
& (spin-triplet) & (mixed-spin) & (spin-singlet) \\ \hline
No Constraint & 11.315 & 7.478 & 8.037 \\\hline
 No ${S=0}$& 11.315 & 6.299 & 1.630 \\\hline
No ${S=1}$ & 4.853 & 4.630 & 8.035 \\\hline 
\end{tabular}
\caption{The correlation energy $E_{\text{corr}}$[MeV]. This quantity is defined as the difference in HFB ground 
state binding energy between the unpaired nucleus and the one subject to pairing 
constraints  (or completely unconstrained).}
\label{table:Ecorr}
\end{table}
As can be seen from Table \ref{table:Ecorr}, the unconstrained ground state and the one found by removing spin-singlet pairing 
are nearly degenerate in energy for ${}^{132}_{66}$Dy. This is an indication of this nucleus exhibiting mainly 
spin-triplet pairings. 
The situation in the case of ${}^{132}_{64}$Gd is quite different. Neither pairing channel is suppressed in the ground state,
and both pair constrained states have very similar values of correlation energy. 
We, thus, can expect ${}^{132}_{64}$Gd to be of spin-mixed pairing nature.
The last isotope, ${}^{132}_{60}$Nd, is analogous to ${}^{132}_{66}$Dy, but for spin-singlet pairing.
Note that when we refer to a nucleus as exhibiting, say, spin-singlet pairing, we merely mean that the
channel is dominant (not that it is the only one present).

\subsection{The eigenbasis}
\label{subsec:basis}
The basis chosen is block diagonal in orbital angular momentum $\hat{L}$ (multiple $l$ values are present),
and diagonal in isospin $\hat{T}$, and spin $\hat{S}$. 
The symmetries we are studying are particle number $\hat{A}$, more specifically neutron and proton number, and nuclear spin ($\hat{\boldsymbol{J}}=\hat{\boldsymbol{L}}+\hat{\boldsymbol{S}}$).
The respective operators are represented by matrices in this basis,
\begin{equation}
 \begin{split}
 \hat{A}^{(LTS)}=& \mathbb{I}_{N_L} \otimes \mathbb{I}_{2} \otimes \mathbb{I}_2 \\
  \hat{T}^{(LTS)}=& \mathbb{I}_{N_L} \otimes J_{1/2} \otimes \mathbb{I}_2 \\
  \hat{J}^{(LTS)}_z=& \{\oplus J_{L_i} \} \otimes \mathbb{I}_2 \otimes \mathbb{I}_2 + \mathbb{I}_{N_L} \otimes \mathbb{I}_2 \otimes J_{1/2}\\
  N_L =& \sum_i(2L_i+1)\\
 \end{split}
\end{equation}
where $J_D$ is the angular momentum operator in the irreducible diagonal representation in which $D$ is the maximal eigenvalue of $\hat{J}_z$ \cite{Georgi2009}.
$\mathbb{I}_{N_L}$ is the $N_L$ dimensional identity matrix. 

\section{Symmetry Restoration}
Following previous work, we find the ground state through energy minimization and then perform symmetry projection, 
also called projection after variation (PAV)  \cite{Grasso2012,Ripoche2017,Lacroix2012}. The energy minimization
procedure does not respect either $\hat{A}$ or $\hat{J}$ conservation, and these symmetries are restored by projecting
out the eigenstate composition of the wave-function found.

The probability for a quantum number $K$ to be present in the wave-function $| \Phi \rangle$ is given by the formula,
\begin{equation}
 \begin{split}
  \langle \Phi|\hat{P}^{K}| \Phi \rangle=\frac{d_K}{\Omega_0}\int d\Omega \ P^{K}_{II}(\Omega)\ \langle \Phi |\mathcal{P}(\Omega)| \Phi' \rangle
  \label{eq:symmetry}
 \end{split}
\end{equation}
where $P^{K}_{II}(\Omega)$ is a diagonal matrix element of the symmetry group $P$ in representation of dimensionality $d_K$, and $\Omega_0$ is
the volume integral of the group.
The overlap $\langle \Phi |\mathcal{P}(\Omega)| \Phi' \rangle$ is calculated based on the expressions from Ref.~\cite{Bertsch2012}. 
The numerical implementation of the Pfaffian is based on the Parlett-Reid algorithm as shown in Ref.~\cite{Wimmer2012}. 
\subsection{Particle-number projection}
\label{subsection:part_num}
In the case of simultaneous projection of proton and neutron number, the projection operator is,
\begin{equation}
\begin{split}
\hat{P}^{(LTS)}_{NZ}(N_0,Z_0)=&\int_0^{2\pi}\frac{d\varphi_N}{2\pi} e^{-i N_0\varphi_N} \int_0^{2\pi}\frac{d\varphi_Z}{2\pi}e^{-i Z_0\varphi_Z}\\
  &\ \ \times \ e^{i R(\varphi_N,\varphi_Z)}\\
  R(\varphi_N,\varphi_Z)=& \mathbb{I}_{N_L} \otimes \begin{pmatrix} { \varphi_N} & 0 \\ 0 & { \varphi_Z} \end{pmatrix} \otimes \mathbb{I}_2
  \label{eq:pnumber}
\end{split}
\end{equation}
\begin{figure}[ht]
 \subfloat[${}^{132}_{66}$Dy]{\includegraphics[scale=0.35]{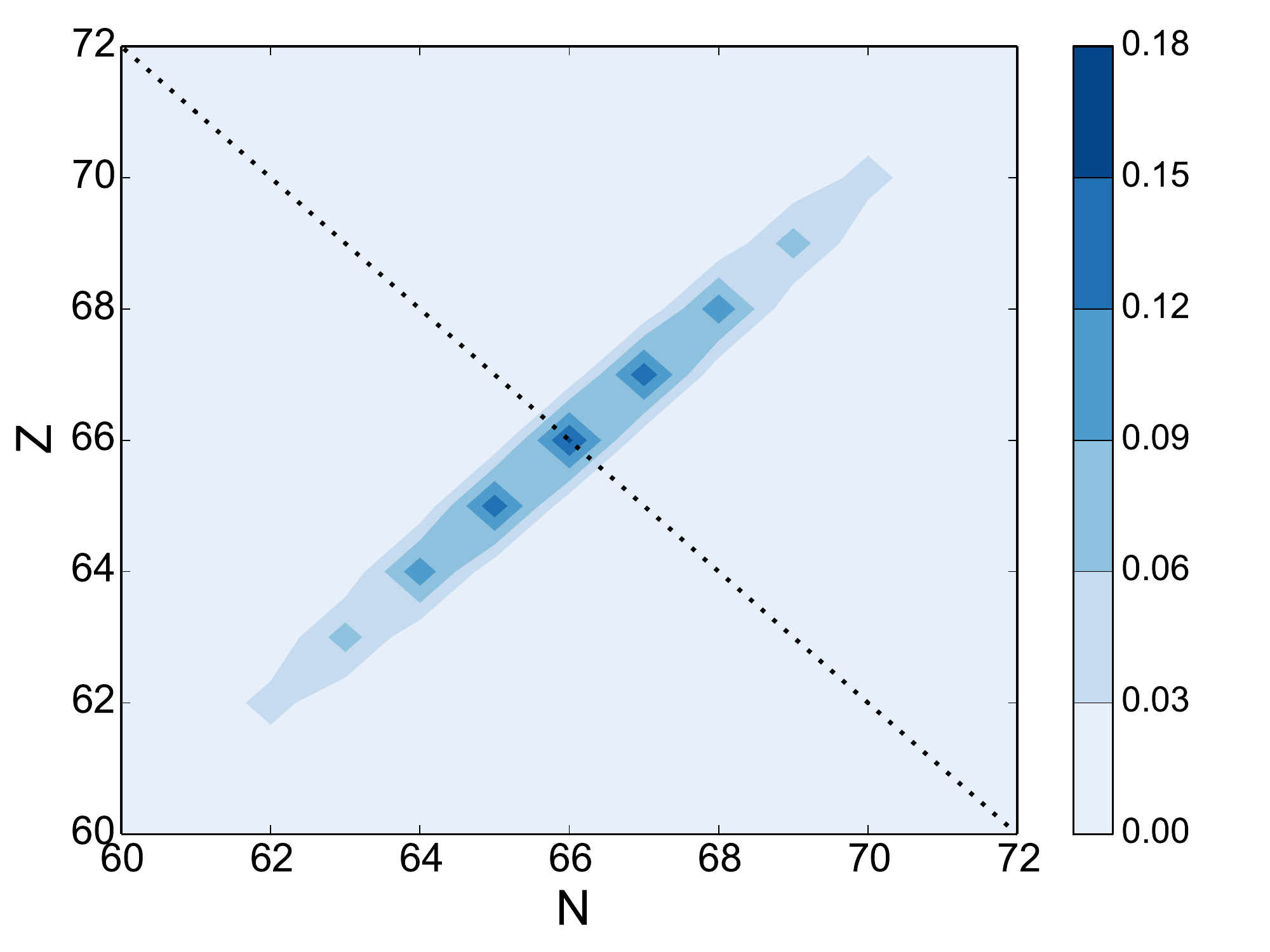}}\\
 \subfloat[${}^{132}_{64}$Gd]{\includegraphics[scale=0.35]{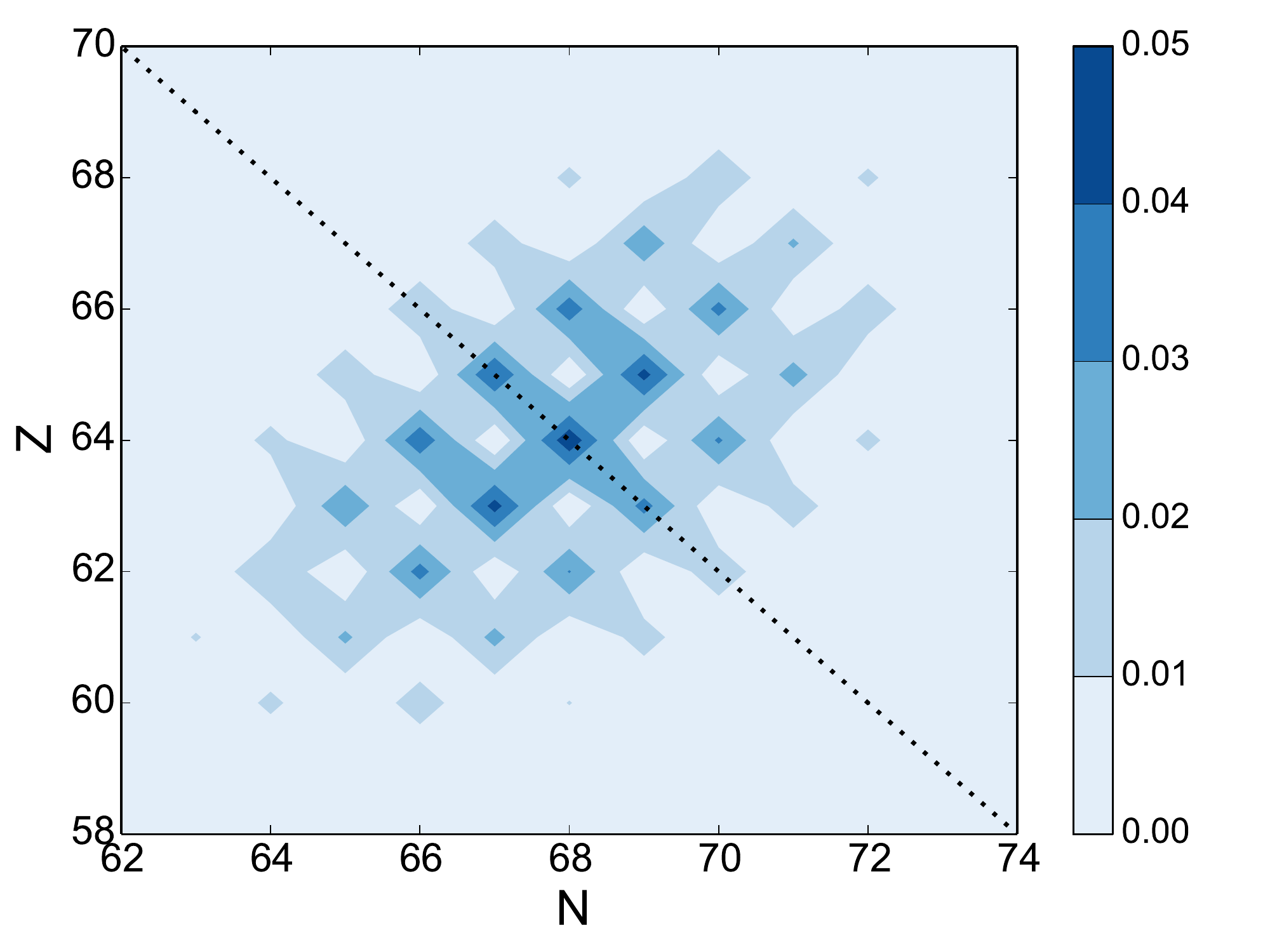}}\\
 \subfloat[${}^{132}_{60}$Nd]{\includegraphics[scale=0.35]{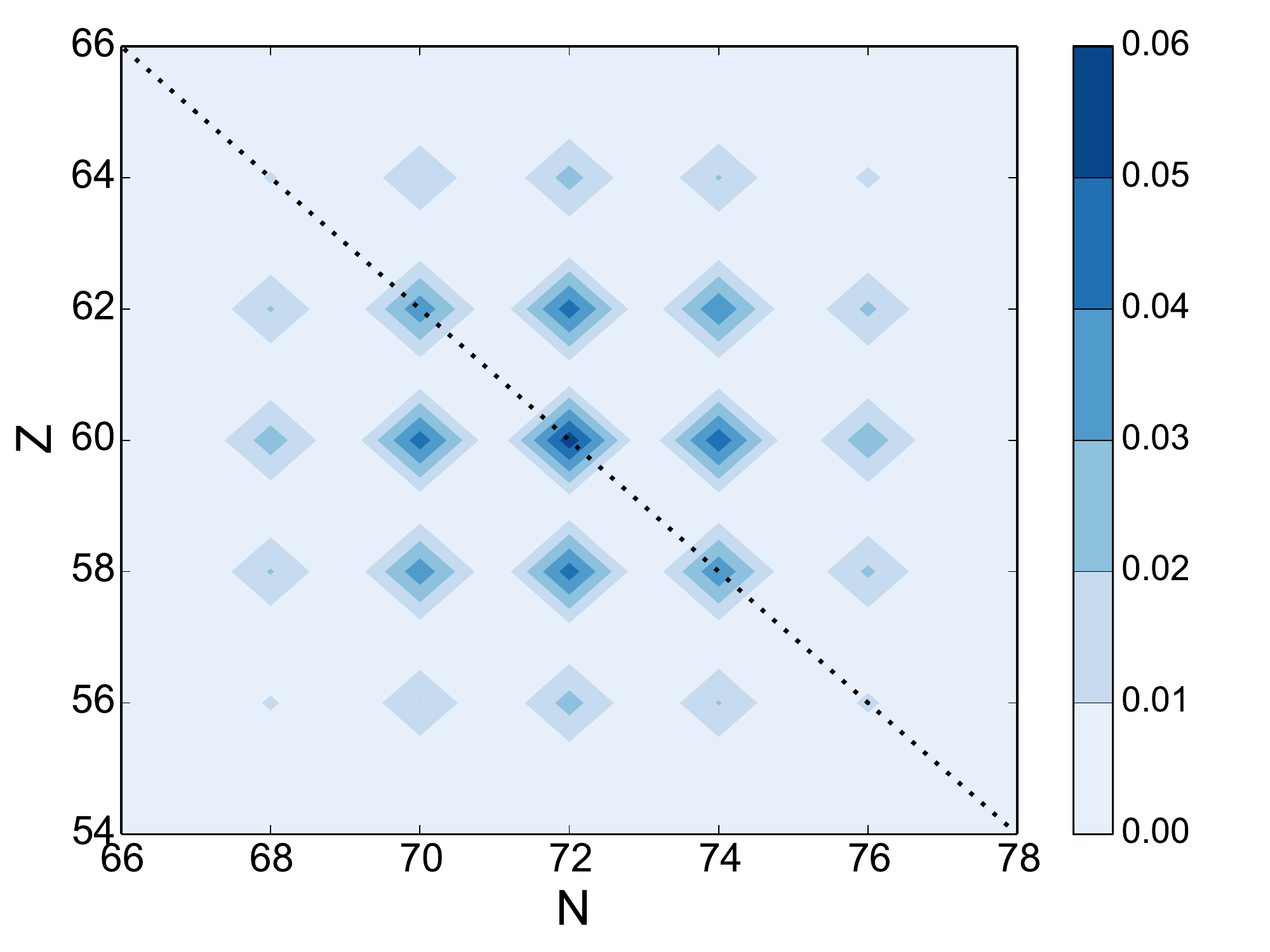}}\\
 \captionsetup{justification=raggedright }
 \caption{Two-dimensional probability distributions,  Eq.~(\ref{eq:symmetry}) with the projection operator in Eq.~(\ref{eq:pnumber}). Dotted line represents $A=132$.}
 \label{fig:2D_no_constraints}
\end{figure}

\begin{figure}[ht]
 \subfloat[No spin-singlet pairing]{\includegraphics[scale=0.35]{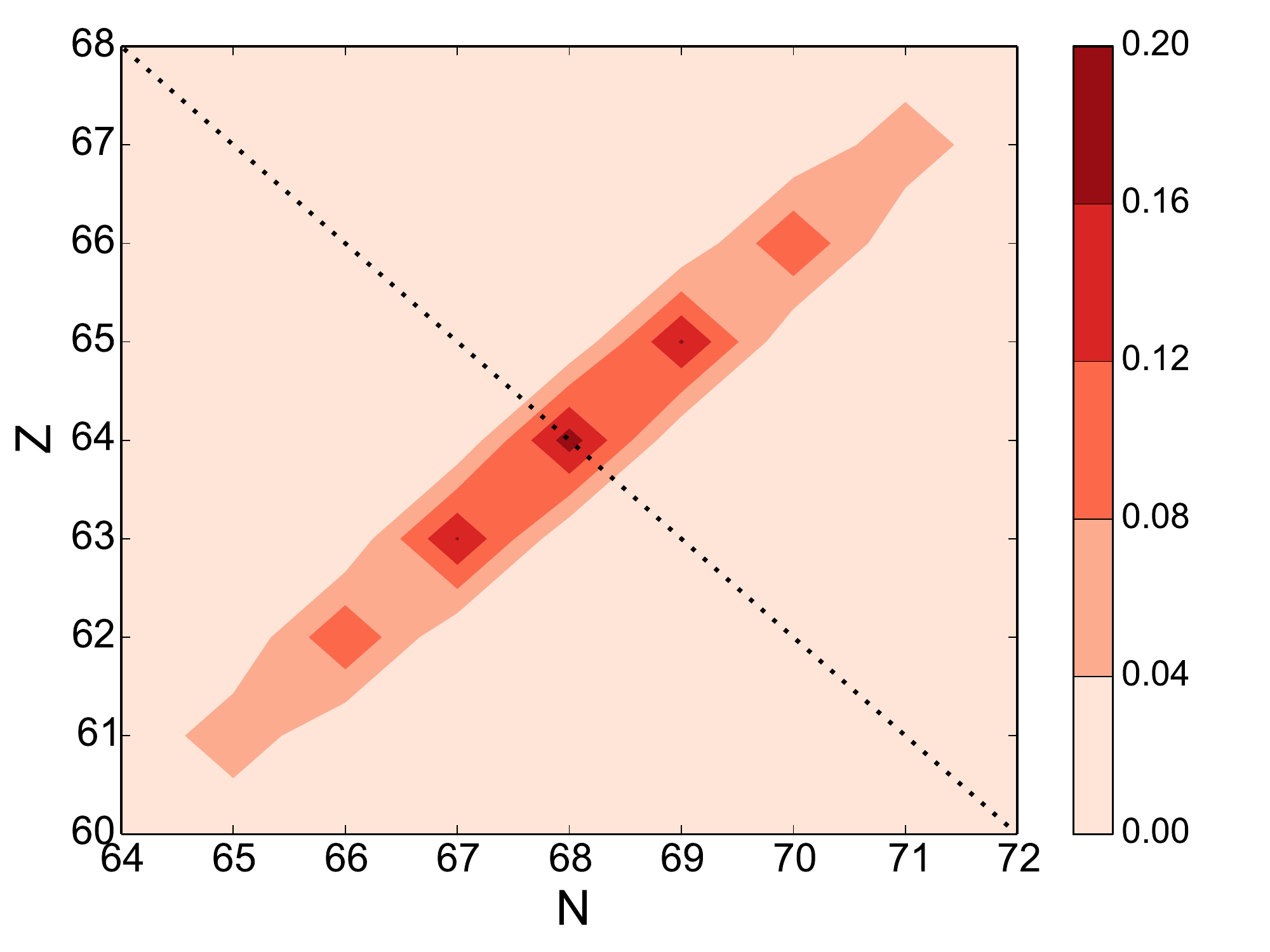}}\\
 \subfloat[No spin-triplet pairing]{\includegraphics[scale=0.35]{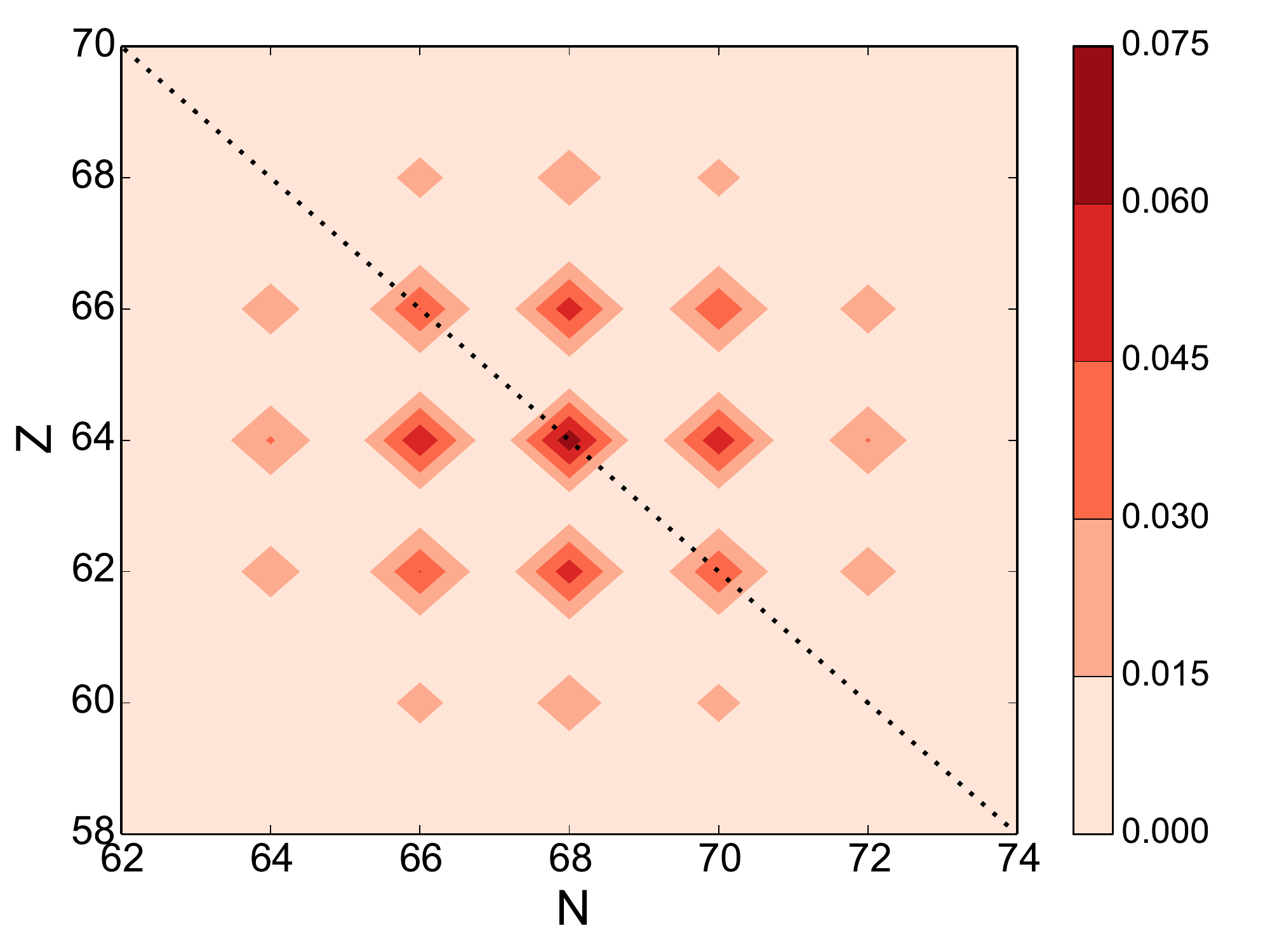}}\\
 \captionsetup{justification=raggedright }
 \caption{Two-dimensional probability distribution for ${}^{132}_{64}$Gd subject to pairing constraints. Dotted line represents $A=132$.}
 \label{fig:2D_Gd}
\end{figure}
As the plot in Fig.~\ref{fig:2D_no_constraints} shows, the presence of only spin-triplet pairing forces the probability
distributions for protons and neutrons to be strongly coupled (the distribution is perpendicular to the $A=132$ line).
If only spin-singlet pairing is present, the distributions are decoupled and there is a checkerboard pattern centered at the target isotope. Mixed-spin pairing
is a hybrid of the two previous configurations.
Note that for each of the three nuclei under study we see contributions coming from 
several even-even and odd-odd nuclides (this is true also for ${}^{132}_{60}$Nd, where the 
odd-odd contributions are tiny but their cumulative contribution to $I_0=1$ is noticeable as shown in the following sections).

To have a better understanding of the pattern observed, it is instructive to perform symmetry restoration on the mixed-spin isotope,
by constraining one type of pairing at a time, to see how the ground state configuration looks in terms of neutron and proton number 
distributions. As displayed in Fig.~\ref{fig:2D_Gd}, the pattern found in Fig.~\ref{fig:2D_no_constraints} persists; spin-triplet pairing 
symmetrizes the distributions while spin-singlet completely decouples them.

As a check, we have carried out further calculations, where 
we remove spin-singlet pairing from the ground state of ${}^{132}_{66}$Dy: 
there was no significant change in the particle number distribution. 
The same turned out to be the case when removing spin-triplet pairing for ${}^{132}_{60}$Nd. 
To avoid any confusion when reading our two-dimensional distribution plots, 
we emphasize here that projection to only integer particle number was performed, 
since $N$ and $Z$ are treated as integers throughout this work.

\subsection{Angular momentum projection (nuclear spin)}
\label{subsection:nuc_spin}
\begin{figure}[ht]
  \includegraphics[scale=0.55]{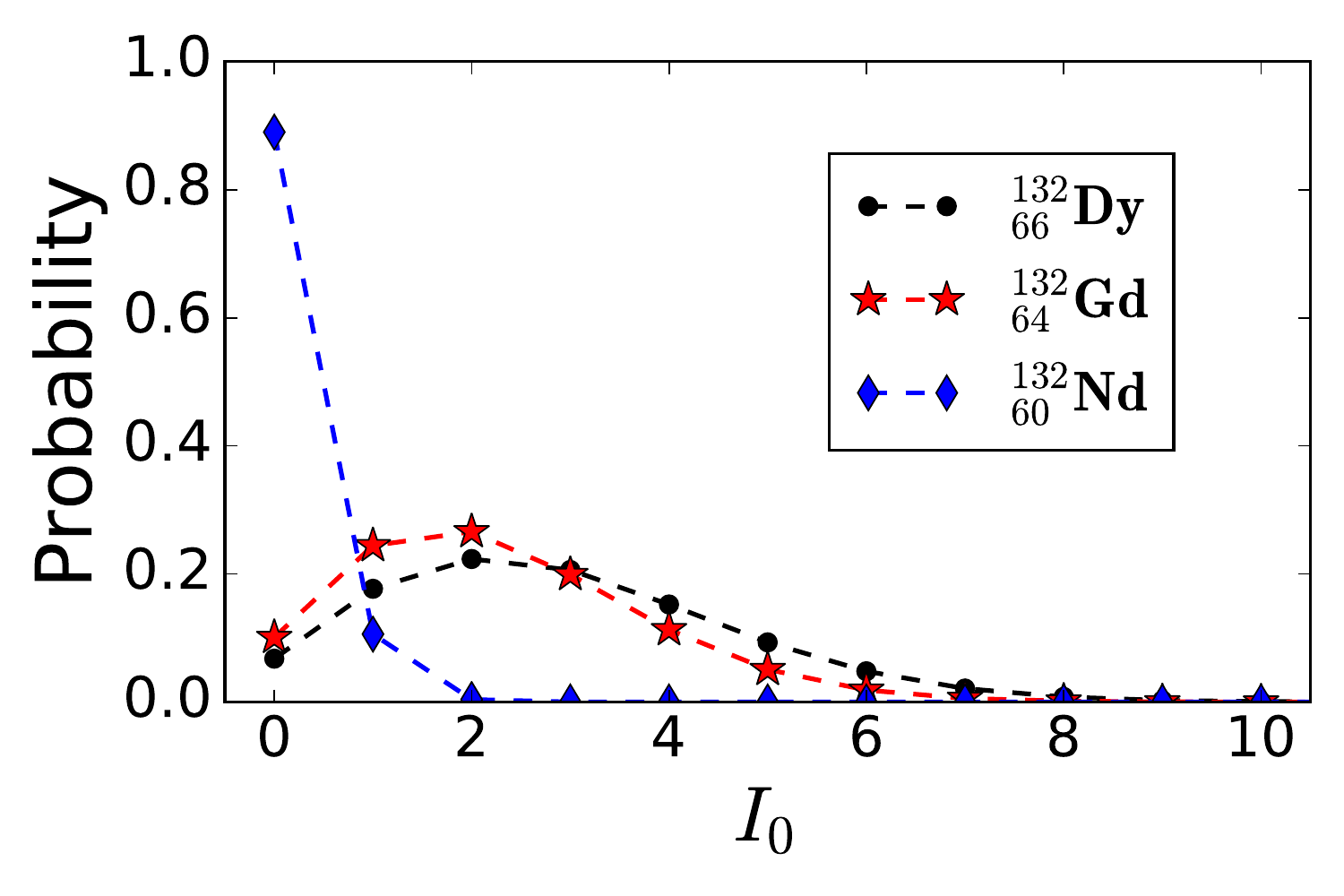}
  \captionsetup{justification=raggedright }
  \caption{$I_0$ probability distribution,  Eq.~(\ref{eq:symmetry}), with the projection operator in Eq.~(\ref{eq:pjz}), for ${}^{132}_{66}$Dy (black circles), ${}^{132}_{64}$Gd (red stars), 
  and ${}^{132}_{60}$Nd (blue diamonds).}
  \label{fig:J_no_constraint}
 \end{figure}
The rotational group is parametrized in terms of the three Euler angles $\Omega=(\alpha,\beta,\gamma)$ and the symmetry group under consideration is SU(2).
The respective expression for $\hat{J}$ projection is \cite{Ring1980}
 \begin{equation}
  \begin{split}
   \hat{P}^{(LTS)}_J(& I_0,m',m)=\frac{2 I_0+1}{8\pi^2}\int_0^{2\pi}d\alpha \int_{0}^{\pi}d\beta\sin(\beta)\\
   &\times \int_0^{2\pi}d\gamma 
   e^{i(m'\alpha+m\gamma)}d^{(I_0)}_{m',m}(\beta)e^{i\alpha \hat{J}^{(LTS)}_z}\\
   & \times \ \ e^{i\beta \hat{J}^{(LTS)}_y}e^{i\gamma \hat{J}^{(LTS)}_z}
  \end{split}
  \label{eq:pjz}
 \end{equation}
where  $d^{(I_0)}_{m', m}(\beta) = \langle I_0, m' | \exp[i \beta \hat{J}_y] | I_0 ,m\rangle$ is
the Wigner matrix representing the rotation matrix element around the $y$ axis in the $\hat{J}_z$ basis \cite{Edmonds1955}. 
We used a slight modification of Eq.~(\ref{eq:pjz}), where the range of integration for $\gamma$ is twice the full rotation,
and the projection operator was normalized accordingly. The reason for this change is to allow for the simultaneous projection of both 
half and full angular momentum values. If the number of quasiparticles is even, only integer values of $I_0$ are to be expected, but for odd-number nuclei, spin-half
values might be present. 

Figure~\ref{fig:J_no_constraint} depicts the $I_0$ probability distribution for all three isotopes. In the case of spin-singlet pairing
(${}^{132}_{60}$Nd) $I_0=0$ is the dominant state, and in the case of spin-triplet pairing (${}^{132}_{66}$Dy), there is a spread peaked at low
values of $I_0$. Interestingly, the mixed-spin paired isotope resembles more the spin-triplet distribution.
 \begin{figure}[b]
  \includegraphics[scale=0.55]{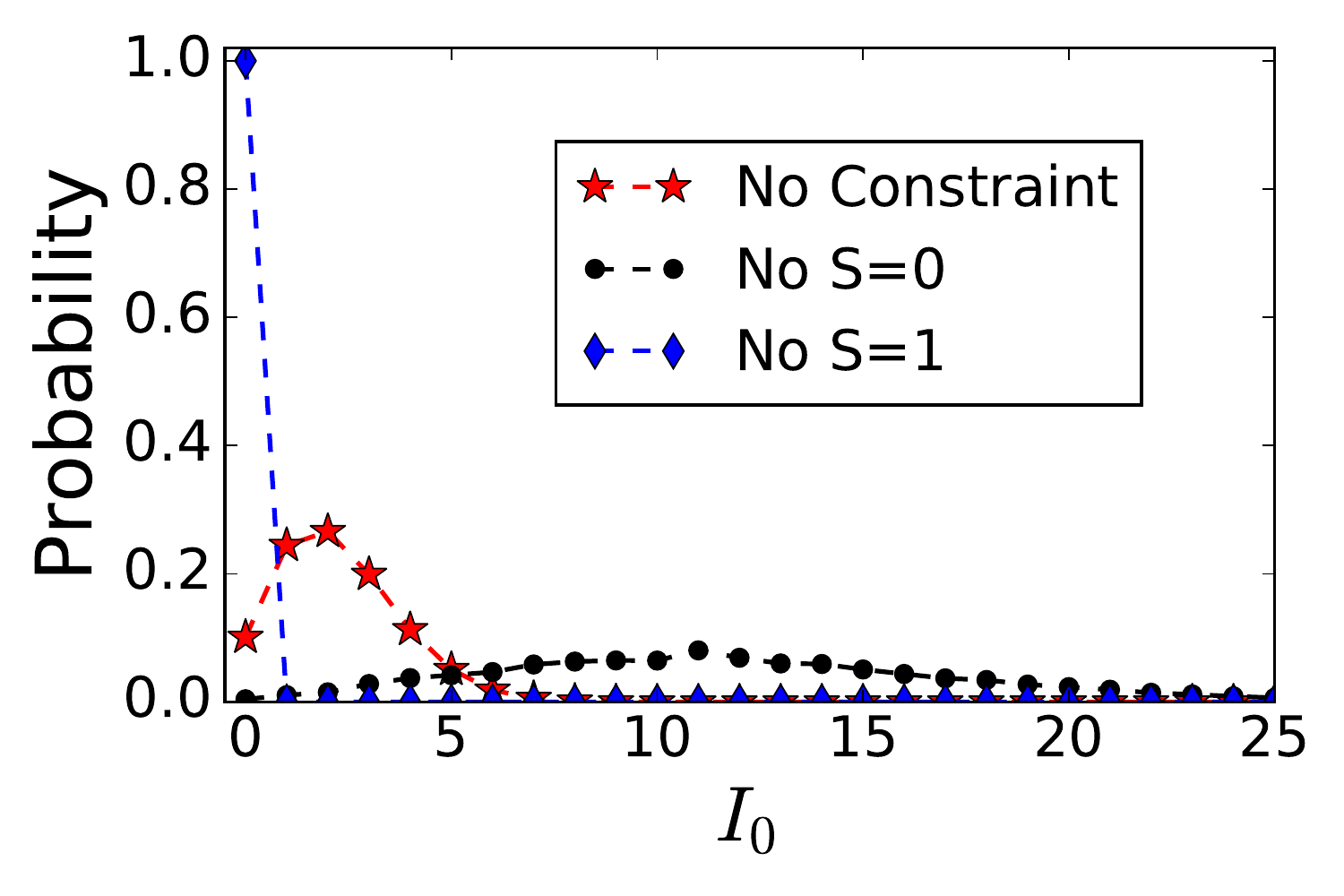}
 \captionsetup{justification=raggedright }
 \caption{$I_0$ probability distribution, as in Fig.~\ref{fig:J_no_constraint}, for ${}^{132}_{64}$Gd subject to no constraint (red stars), no spin-singlet (black circles), 
 and no spin-triplet (blue diamonds) pairing.}
 \label{fig:J1D}
 \end{figure}
The probability distributions for $I_0$ subject to all the pairing constraints for the mixed-spin pairing isotope are depicted in Fig.~\ref{fig:J1D}.
A rather intriguing pattern emerges from this figure;
when only spin-singlet pairing is present, $I_0=0$  is the only value present, and when only 
spin-triplet pairing is present, there is a wide spread of possible $I_0$ values. 
\subsection{Particle number and angular momentum}
\label{subsection:nuc_spin_n}
\begin{figure}[t]
 \subfloat[${}^{132}_{66}$Dy]{\includegraphics[scale=0.45]{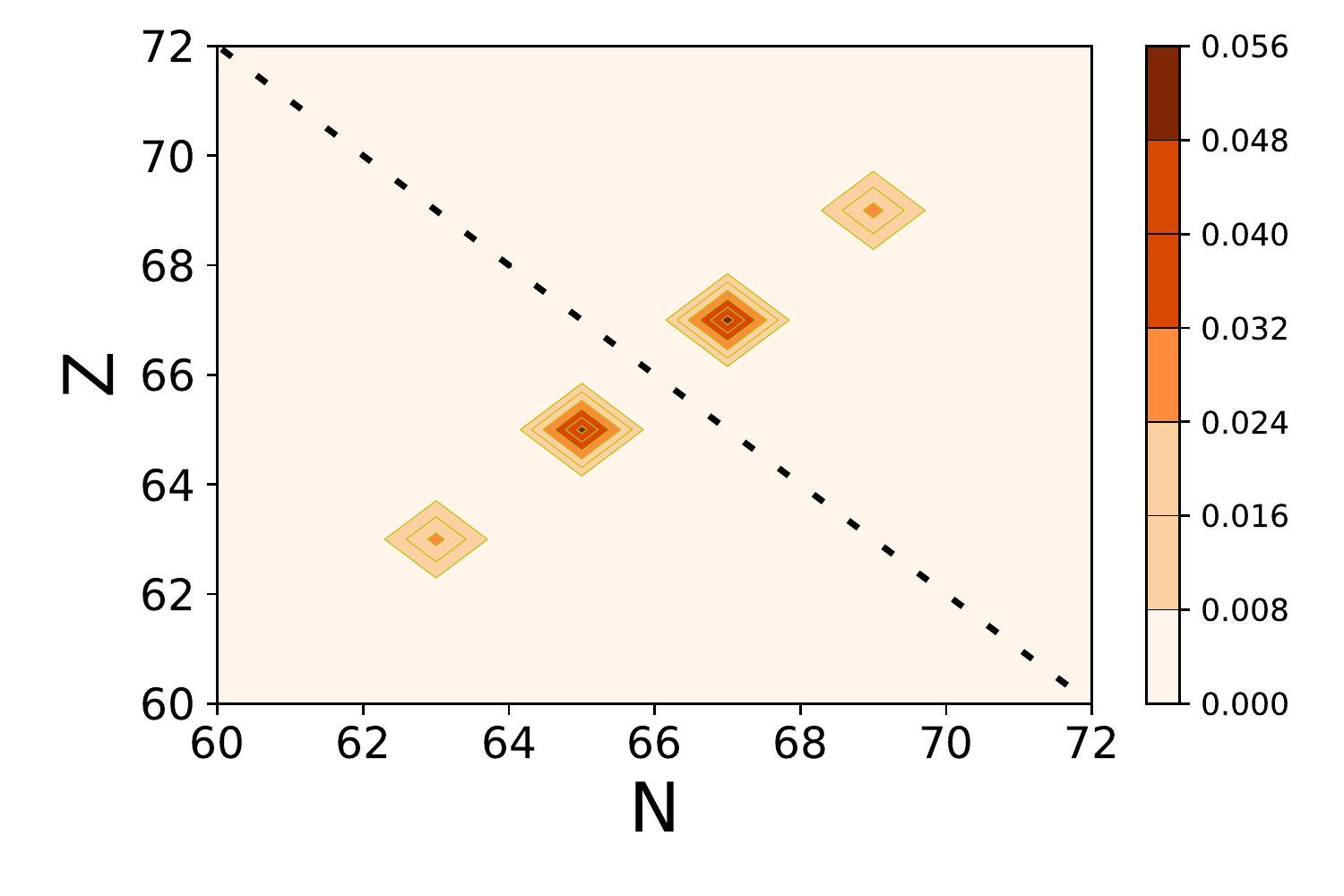}}\\
 \subfloat[${}^{132}_{64}$Gd]{\includegraphics[scale=0.45]{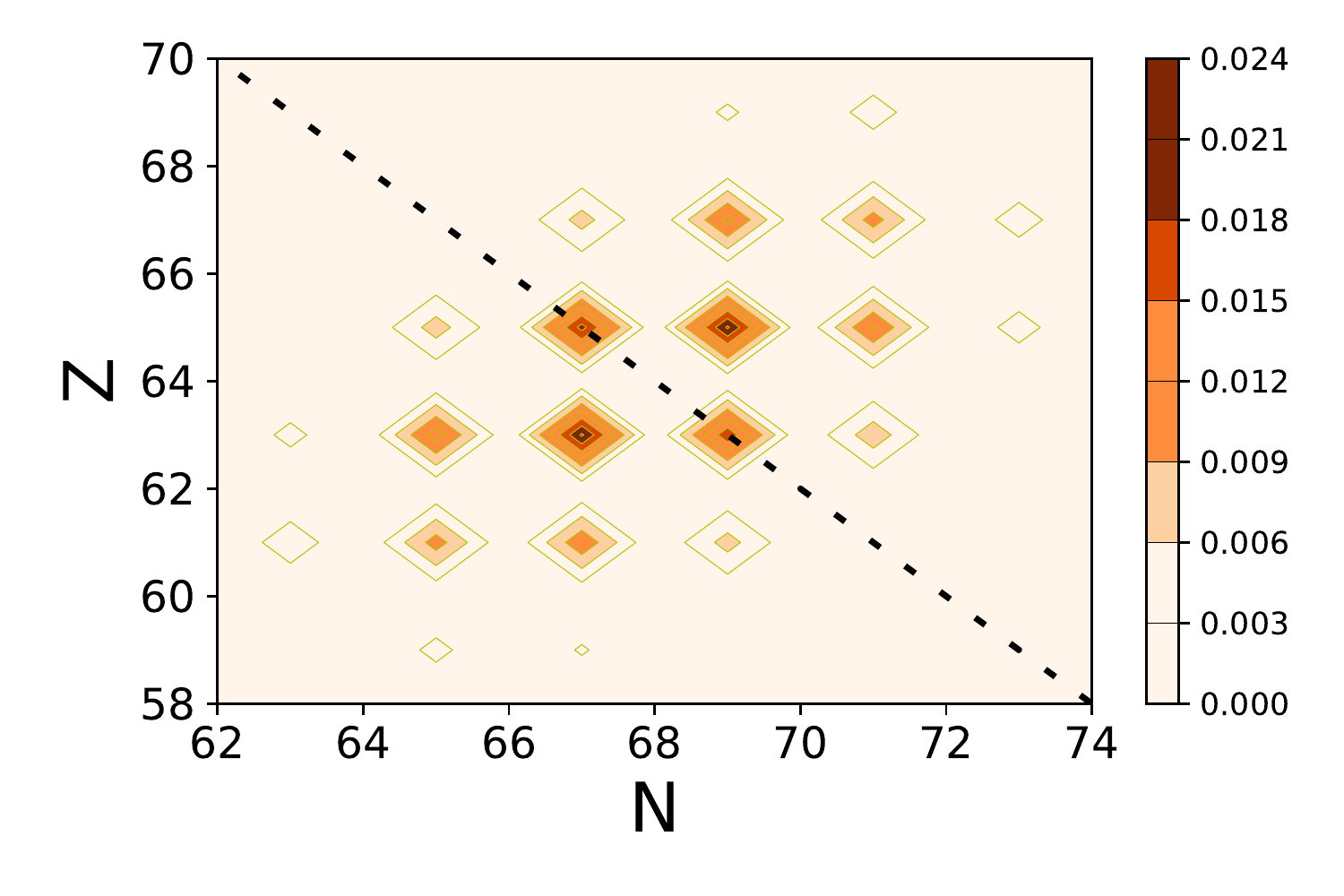}}\\
 \subfloat[${}^{132}_{60}$Nd]{\includegraphics[scale=0.45]{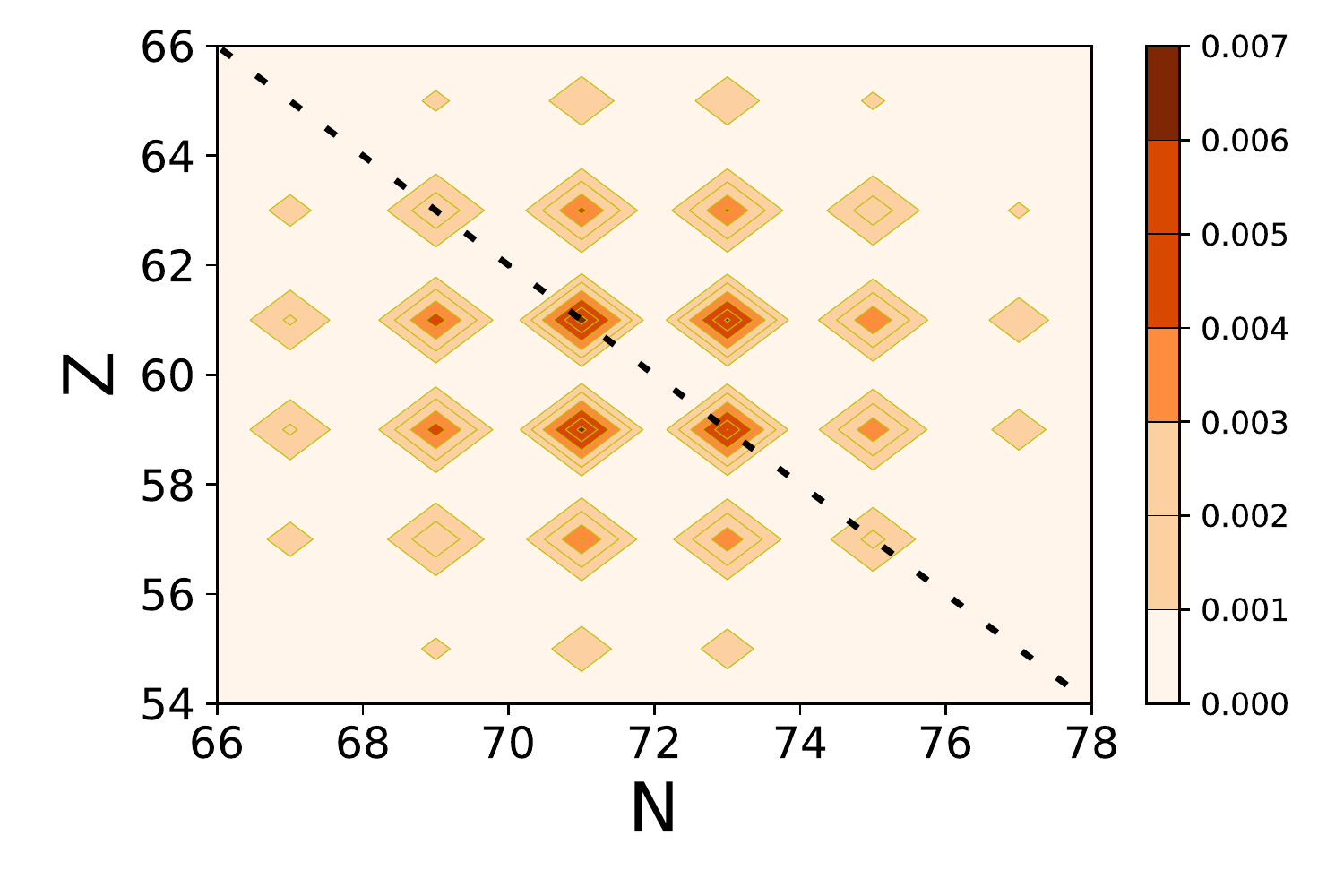}}\\
 \captionsetup{justification=raggedright }
 \caption{Two-dimensional probability distributions for the three isotopes for $I_0=1$,  Eq.~(\ref{eq:symmetry}) with the projection operator in Eq.~(\ref{eq:lts}), without any constraint on pairing. 
 Dotted line represents $A=132$.}
 \label{fig:2D_j1}
\end{figure}

To identify what fraction of the ground state has the ``right'' quantum numbers ($N_0,Z_0,I_0$), 
a simultaneous projection is required:
\begin{equation}
 \begin{split}
  \hat{P}^{(LTS)}_{NZJ}(N_0,Z_0,I_0)=\hat{P}^{(LTS)}_{NZ}(N_0,Z_0)\sum_{m=-I_0}^{I_0}\hat{P}^{(LTS)}_J(I_0,m,m).
  \label{eq:lts}
 \end{split}
\end{equation}
\begin{figure}[b]
 \subfloat[${}^{132}_{66}$Dy]{\includegraphics[scale=0.3]{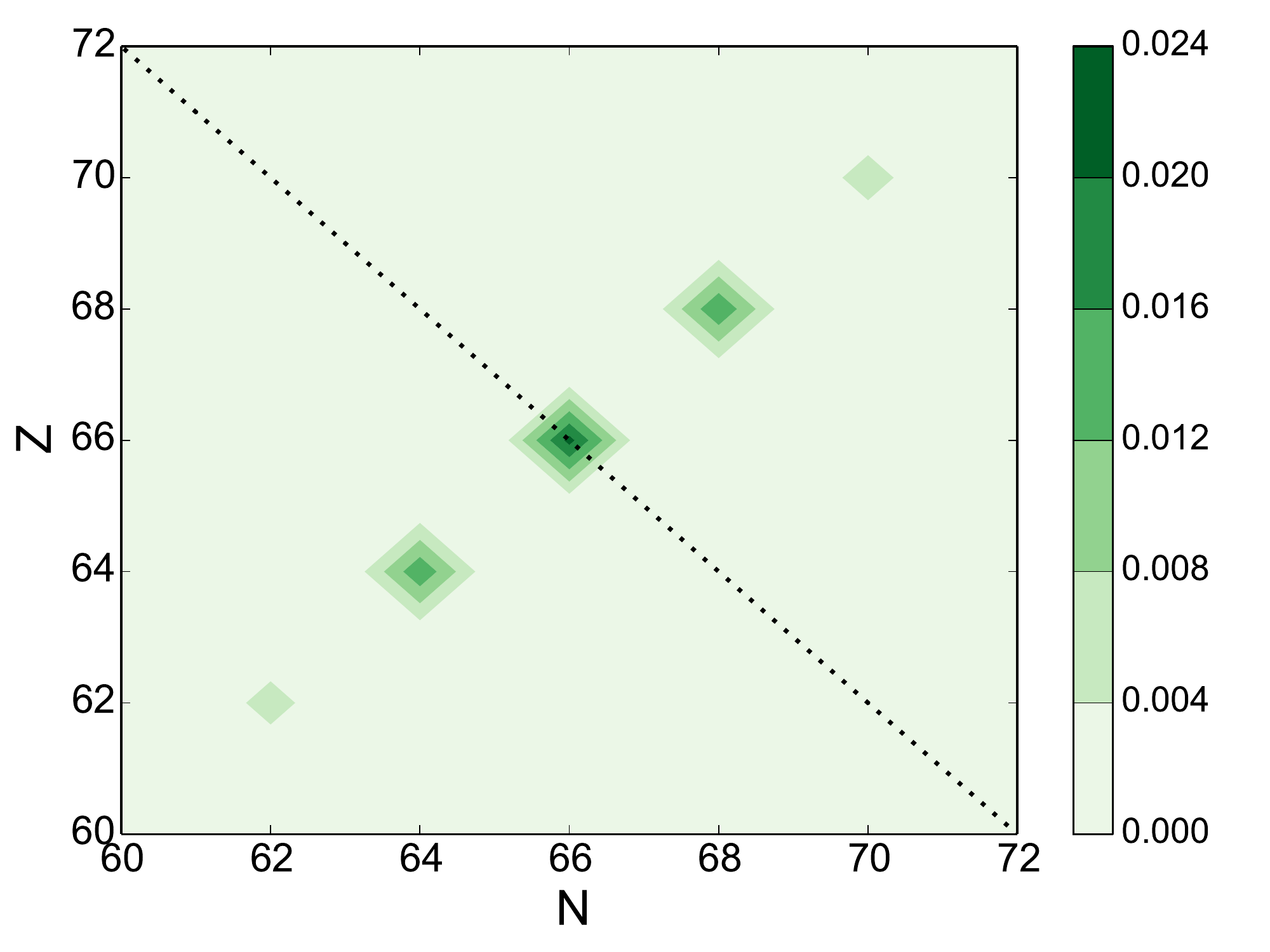}}\\
 \subfloat[${}^{132}_{64}$Gd]{\includegraphics[scale=0.3]{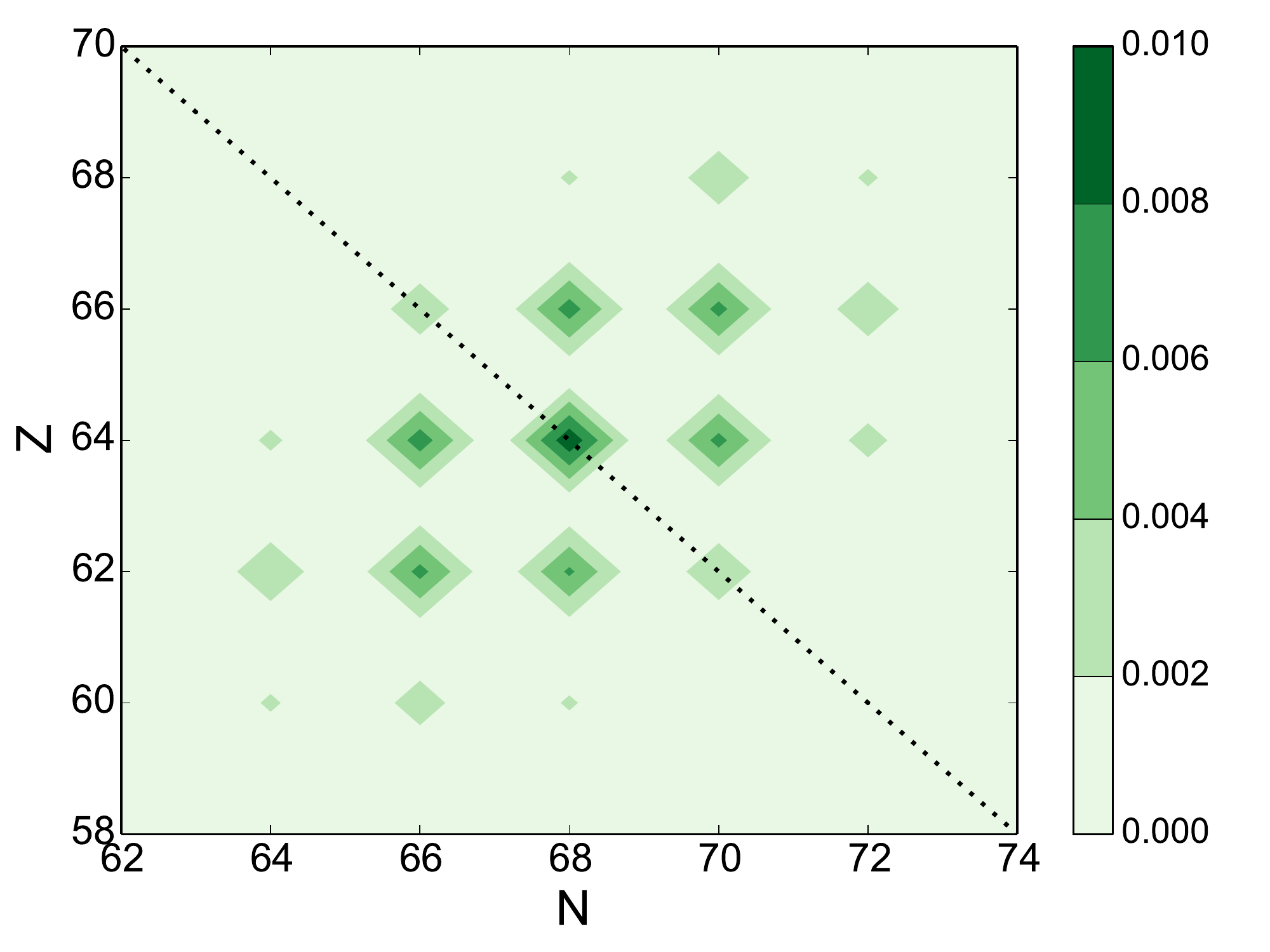}}\\
 \subfloat[${}^{132}_{60}$Nd]{\includegraphics[scale=0.3]{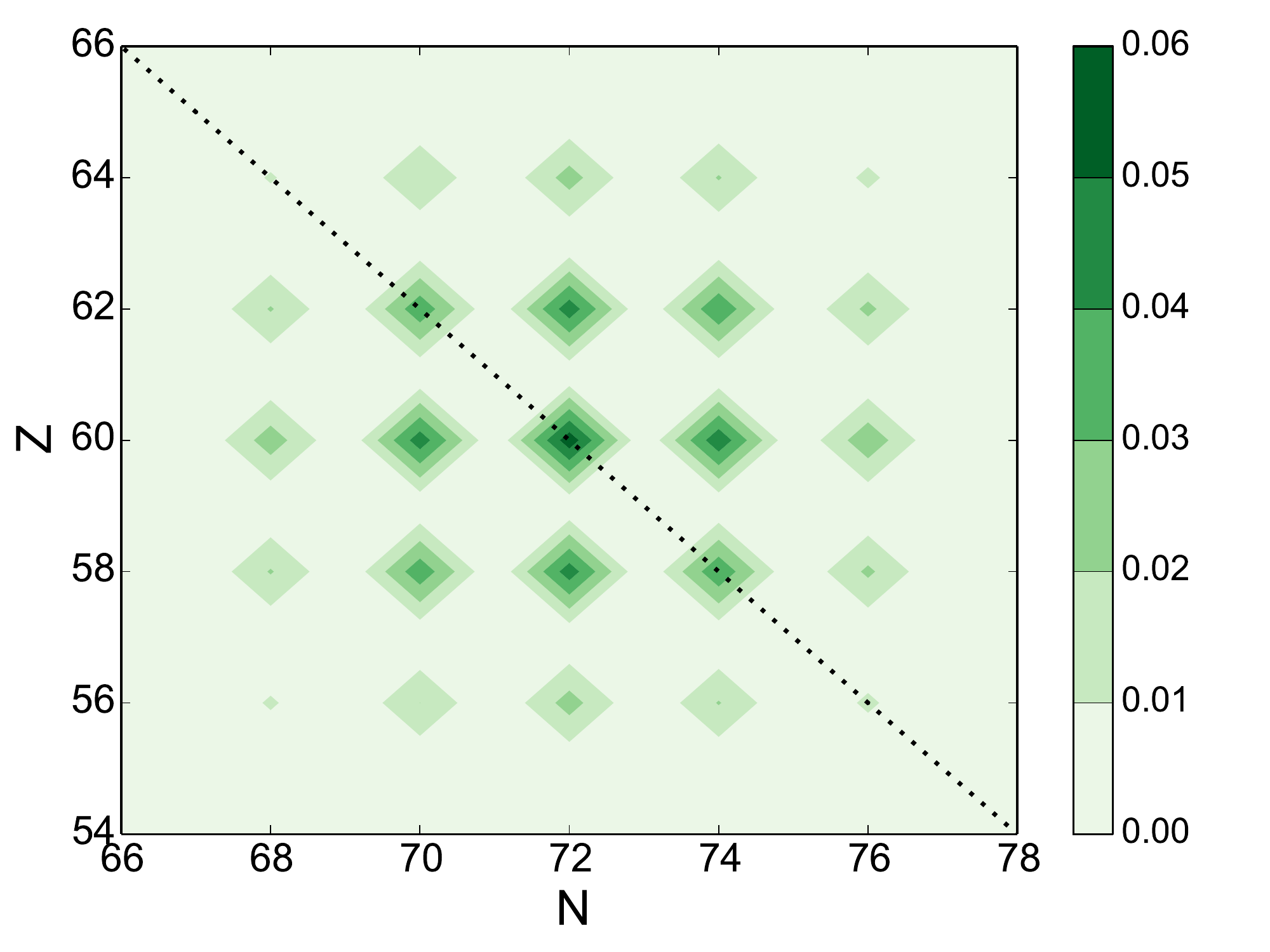}}\\
 \captionsetup{justification=raggedright }
 \caption{Two dimensional probability distributions for the three isotopes for $I_0=0$,  Eq.~(\ref{eq:symmetry}) with the projection operator in Eq.~(\ref{eq:lts}), without any constraint on pairing. 
 Dotted line represents $A=132$.}
 \label{fig:2D_j0}
\end{figure}
In Fig.~\ref{fig:2D_j0} we plot the particle-number probability distributions for the three isotopes after we have projected to $I_0=0$ 
chosen to represent the ground state. 
As can be seen from the plot, for ${}^{132}_{66}$Dy and ${}^{132}_{64}$Gd there is rather a sparse probability distribution which agrees with the result
of Fig.~\ref{fig:J_no_constraint} which shows that $I_0=0$ is a very small part of the wave function. 
The distribution for ${}^{132}_{60}$Nd is almost the same as in Fig.~\ref{fig:2D_no_constraints}, as 90 \% of the wave function has $I_0=0$.
A rather interesting feature emerges from this figure; there are no odd-odd nuclei 
making up the distribution for $I_0=0$, despite the fact that they were
present present in each of the full HFB ground states (Fig. \ref{fig:2D_no_constraints}).  To further understand this situation, 
we also carried out separate calculations, projecting to $I_0=1$ and examining 
the particle-number distribution for each isotope. The result of the projection is depicted in Fig.~\ref{fig:2D_j1}.
We 
(correspondingly) find that only odd-odd nuclei make up the $I_0=1$ distributions.

\begin{figure}[t]
  \includegraphics[scale=0.55]{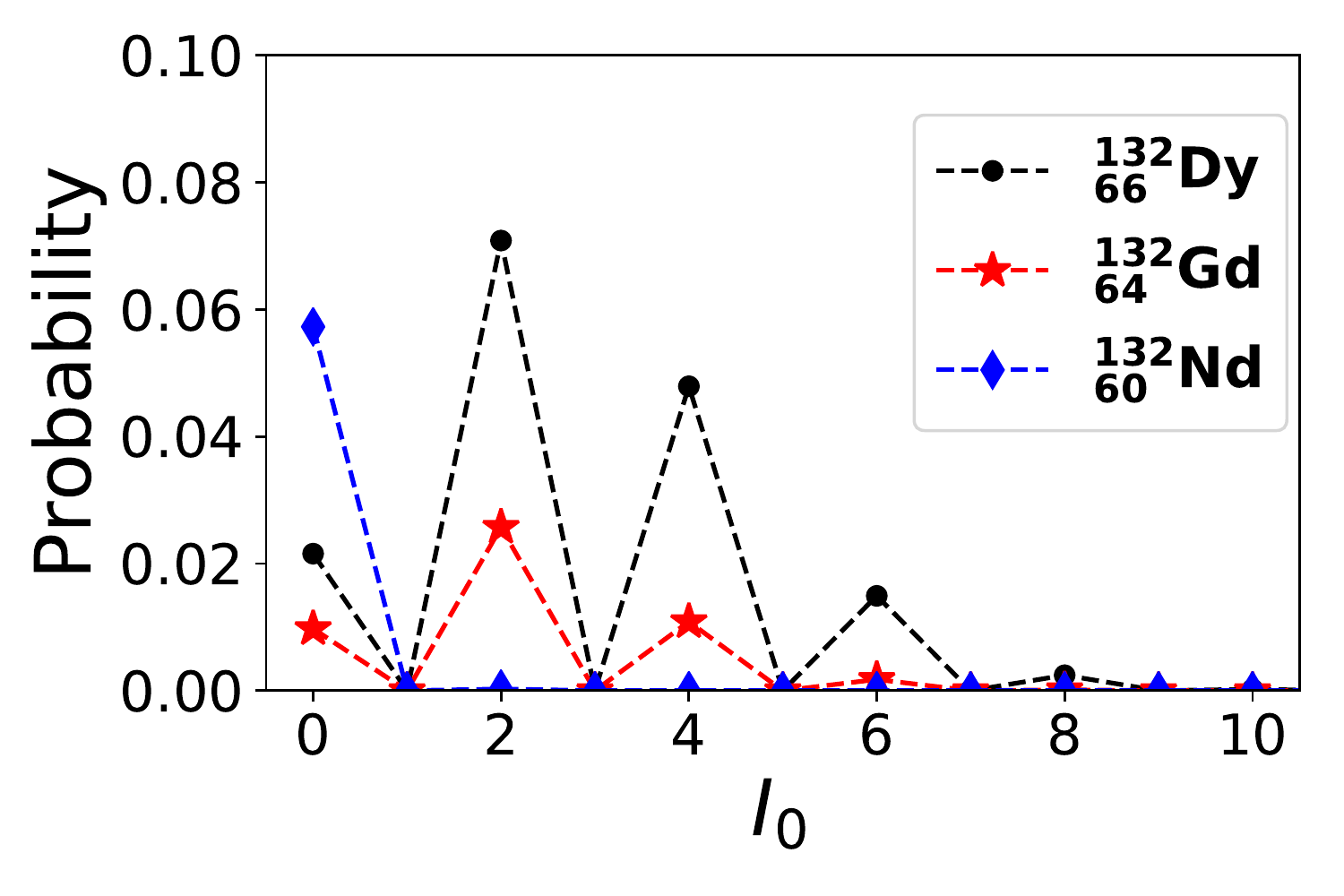}
  \captionsetup{justification=raggedright }
  \caption{$I_0$ probability distribution for ${}^{132}_{66}$Dy (black circles), ${}^{132}_{64}$Gd (red stars), 
  and ${}^{132}_{60}$Nd (blue diamonds). Each isotope has been projected to the respective target neutron and proton numbers.}
  \label{fig:J_nz}
 \end{figure}
 Figure~\ref{fig:J_nz} depicts the $I_0$ distribution for each isotope after the neutron and proton particle numbers have been projected to
 the target values. By comparing with Fig.~\ref{fig:J_no_constraint}, we notice that ${}^{132}_{60}$Nd has only $I_0=0$ for the target particle numbers,
 while the two other isotopes have the same qualitative shape as in the previous plot.
 
\section{Pair Transfer}
\label{section:emission}

\subsection{Wave-function overlap and particle creation operators}
Apart from analyzing the eigen-composition of the HFB ground state, we are also interested in
applying symmetry restoration to  ground-state to ground-state pair-transfer reactions. While this overlap has been treated extensively
in various approximations \cite{Grasso2012} \cite{Shimoyama2011}, our focus here is in finding the most probable pair-transfer 
reaction for nuclei where spin-triplet pairing could be present. 
In particular, we study all the overlaps between the three nuclei under study and the neighboring isotopes 
that can be reached by the addition of two nucleons. 
Instead of assuming that the initial and final nuclei are the same, as is sometimes done, we explicitly include
the appropriate HFB nuclei.
The expressions for the overlap with inclusion of addition or removal of particles are derived from Ref.~\cite{Bertsch2012},
\begin{equation}
 \begin{split}
  &\langle \Phi|\mathcal{P}(\Omega)c^{\dagger}_{q_1}c^{\dagger}_{q_2}| \Phi '\rangle = \frac{(-1)^{n(n-1)/2}}{\langle \Phi | \Phi '\rangle}\\
  & \times \ \text{pf}
  \begin{pmatrix} {V}^t {U} &&  V^t P^{*} q^t  && {V}^t {P}^{*} {V}'{}^{*} \\ -q P^{\dagger} V && 0 && 0 \\
  - {V}'{}^{\dagger} {P}^{\dagger}\ {V} && 0 && {U}'{}^{\dagger} {V}'{}^{*}\end{pmatrix}\\
 \end{split}
\end{equation}
where the($U$, $V$) matrices, describing the wave-function in Eq.~(\ref{eq:wvfn}), have dimensions ($2 n,\ 2 n$), $q$ is a ($2, 2n$) matrix whose rows are the vector representations of the particles
creation operators in the wave-function basis.
Note that, $\langle \Phi|\mathcal{P}(\Omega)c^{\dagger}_{q_1} c^{\dagger}_{q_2}| \Phi '\rangle= \langle \Phi'|c_{q_1} c_{q_2}\mathcal{P}^{\dagger}(\Omega)| \Phi \rangle^{*}$,
so the expression provided can be used for both pair addition or removal.

\subsection{Creation-operator representation}
\label{subsec:create}
The projection operator and its representation has been dealt with in Sec. \ref{subsec:basis}. Here, we describe how to construct the creation
operators of specific quantum numbers ($I_0$,$m_I$,$m_T$).

We start with a basis diagonal in ($\bm{\hat{J}}{}^2\ \hat{J}_z$),
where we assume that also ($\hat{J}_x,\ \hat{J}_y$) are in their standard representation \cite{Georgi2009}.
A creation operator of specific ($I_0,\ m_I$) quantum numbers is represented by
$\bm{e}_i$, the $i$th column of the identity matrix $\mathbb{I}$,
which is also an eigenvector of $\hat{J}_z$, 
\begin{equation}
  \begin{split}
   \bm{q}=\bm{e}_i,\ \bm{\hat{J}}{}^2\bm{e}_i=I_0(I_0+1)\bm{e}_i,\ \hat{J}_z\bm{e}_i=m_I\bm{e}_i
  \end{split}
 \end{equation}
  Given that ($\bm{\hat{J}}{}^2,\ \hat{J}_z$) commute, the transformation from the ($\bm{\hat{L}},\bm{\hat{S}}$) basis to the $\hat{J}_z$ basis is achieved through constructing a matrix pencil  \cite{Golub1996}.
An additional similarity transformation which sets ($\hat{J}_x,\ \hat{J}_y$) in their standard forms and
leaves ($\bm{\hat{J}}{}^2,\ \hat{J}_z$) invariant is required. Let us denote the successive application of these two transformations as 
$Q$:
\begin{equation}
 \begin{split}
  J_z^{(LS)} = Q \ J_z^{(j,m_j)} \ Q^{\dagger}  
 \end{split}
\end{equation}
And, in the basis of the HFB wavefunction,
\begin{equation}
 \begin{split}
  \{\oplus J_{L_i}\} \otimes \mathbb{I}_2 +  \mathbb{I}_{N_L} \otimes J_{1/2}=J_z^{(LS)}
 \end{split}
\end{equation}
The order of the operators in the HFB basis is orbital angular momentum---isospin---spin ($LTS$), and we need to have isospin---orbital angular momentum---spin ($TLS$).
The order reshuffling can be performed with the use of permutation matrices $S_{p,r}=\sum_{i=1}^{p} \bm{e}^t_i \otimes \mathbb{I}_r \otimes \bm{e}_i$ \cite{Harold1981}.
The main property of these matrices is to change the order of a Kronecker product.
The complete reordering between the two bases is performed,
\begin{equation}
 \begin{split}
  J_z^{(LTS)}=&\{\oplus J_{L_i}\} \otimes \mathbb{I}_{2} \otimes \mathbb{I}_{2} + \mathbb{I}_{N_L} \otimes \mathbb{I}_{2} \otimes J_{1/2}\\ 
  =&S_{N_L,4}  \ \bigg[ \mathbb{I}_{2} \otimes \bigg(\mathbb{I}_{2} \otimes \{ \oplus J_{L_i} \}+ J_{1/2} \otimes \mathbb{I}_{N_L} \bigg)   \ \bigg] S_{N_L,4}^t\\
  =&S_{N_L,4}  \ \bigg[ \mathbb{I}_{2} \otimes \bigg(S_{2,N_L}\ J_z^{(LS)}\ S^t_{2,N_L}\bigg) \ \bigg] S_{N_L,4}^t\\
  \end{split}
  \end {equation}
As the careful reader might notice, two successive permutations are performed, the first one is ($LTS \longrightarrow TSL$) and the second one is ($TSL \longrightarrow TLS$). 
This leads us to connect the basis used to find the HFB ground state
with a basis in which particle creation or annihilation operators with specific nuclear spin quantum numbers can be easily expressed in matrix notation.
\begin{equation}
 \begin{split}
  \bm{q}{}^{(\text{HFB basis})} = S_{N_L,4}  \ \bigg[ \mathbb{I}_{2} \otimes \bigg(   S_{2,N_L} \ Q \ \bm{e}_i \ Q^{\dagger} \ S^t_{2,N_L} \bigg)\bigg] \ S^t_{N_L,4}
 \end{split}
\end{equation}

\subsection{Pair-transfer amplitude}
\begin{figure}[ht]
 \subfloat[Final isotope is ${}^{132}_{66}$Dy.]{\includegraphics[width=0.9825\linewidth]{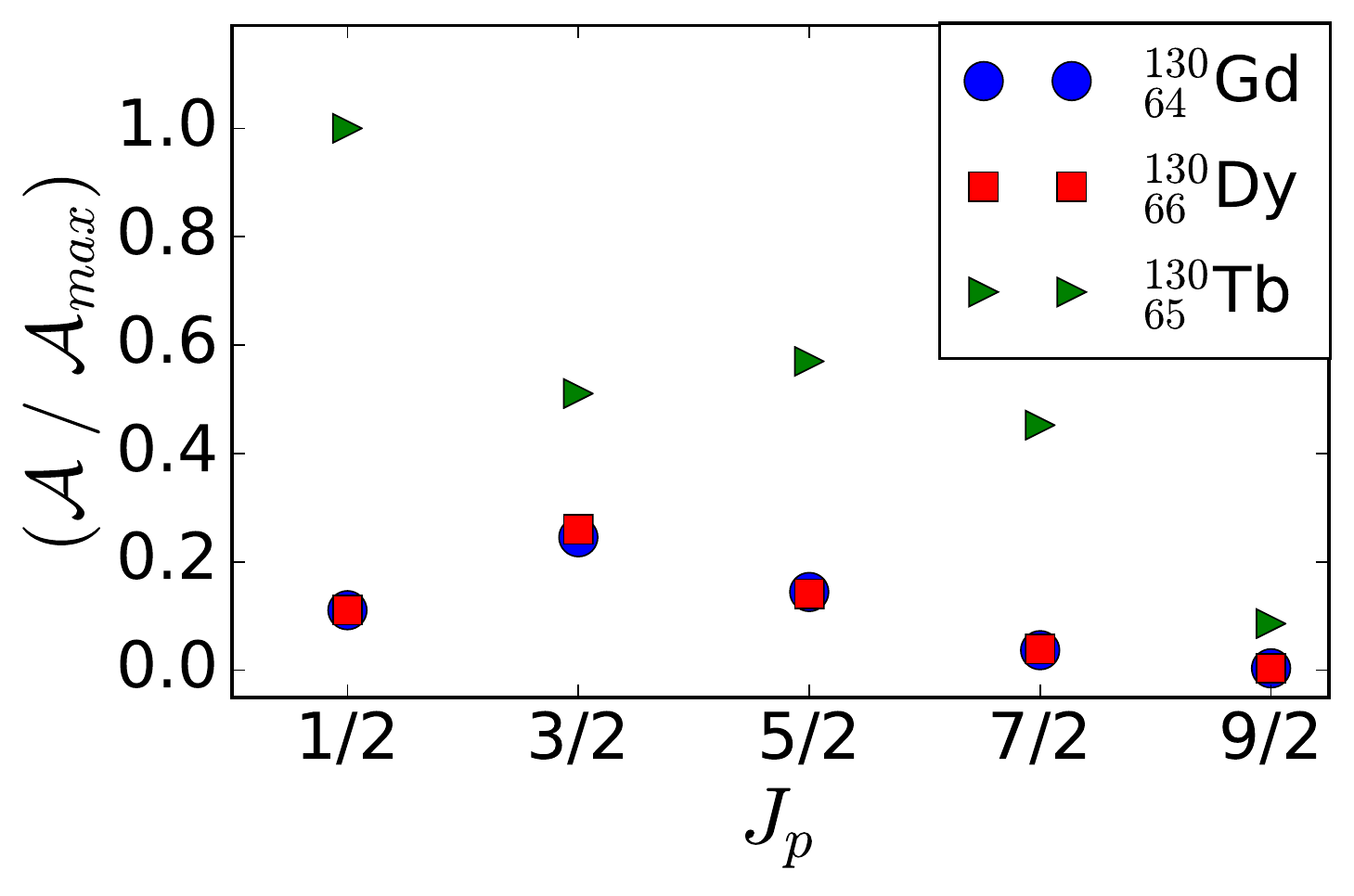}}\\
 \subfloat[Final isotope is ${}^{132}_{64}$Gd.]{\includegraphics[width=0.9825\linewidth]{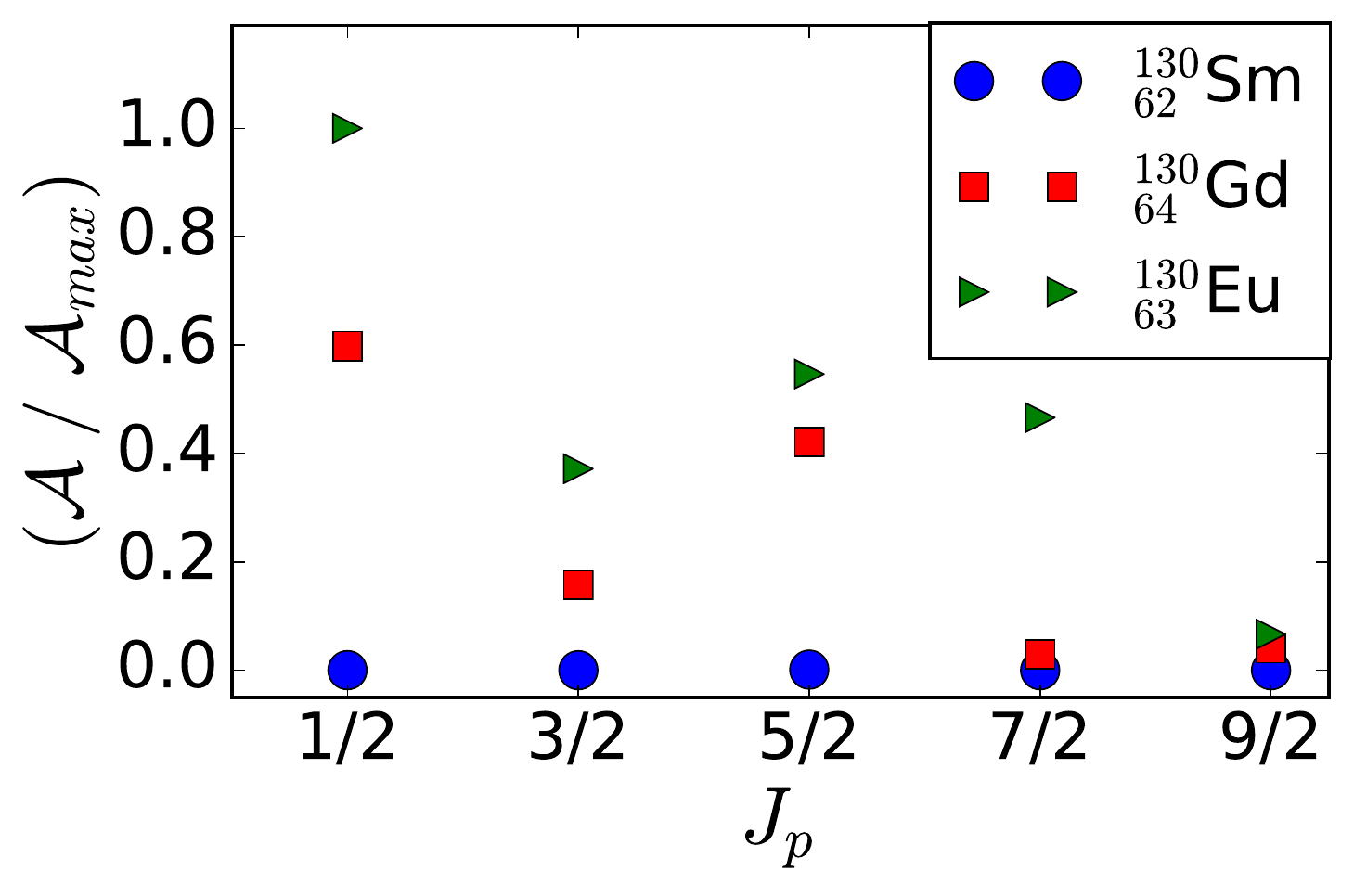}}\\
 \subfloat[Final isotope is ${}^{132}_{60}$Nd.]{\includegraphics[width=0.9825\linewidth]{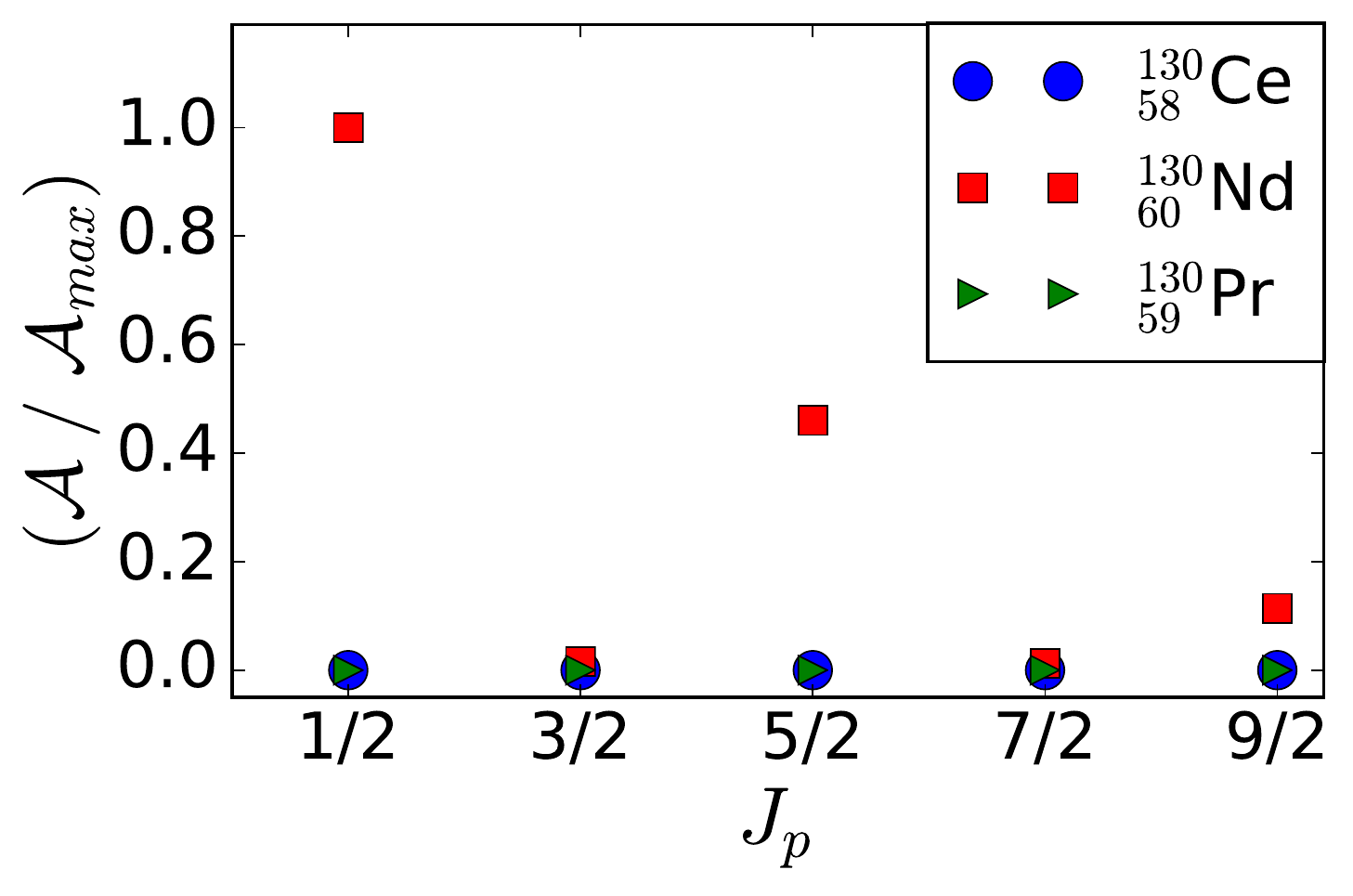}}\\
 \captionsetup{justification=raggedright }
 \caption{Transfer amplitudes for pair addition processes. Blue circles correspond to $pp$ transfer; red squares correspond to
     $nn$ transfer, and green triangles correspond to $np$ pair transfer.}
 \label{fig:transfer}
\end{figure}

To estimate which pair-transfer reaction is more probable, for each of the three nuclei studied so far,
we define the pair-transfer amplitude rate as follows:
\begin{equation}
 \begin{split}
  &\mathcal{A}_{\Phi_i,\Phi_f}^{(J_p)}(I_i,I_f)=|\frac{\langle \Phi_f|\hat{\mathscr{P}}(I_f,J_p)|\Phi_i \rangle}
  {\mathcal{N}_i\mathcal{N}_f}|\\
  &\hat{\mathscr{P}}(I_f,J_p)=\sum_{m_{j_p}=-J_p}^{J_p} \hat{P}_{J}{(I_f)} 
  \hat{c}^{\dagger}{}^{(J_p,-m_{j_p})} \hat{c}^{\dagger}{}^{(J_p,m_{j_p})}\\
  &\mathcal{N}_i=\sqrt{\langle \Phi_i|\hat{P}_{J}{(I_i)}|\Phi_i \rangle}; \ \mathcal{N}_f=\sqrt{\langle \Phi_f|\hat{P}_{J}{(I_f)}|\Phi_f \rangle}
 \label{eq:transfer}
 \end{split}
\end{equation}
where ($I_i,I_f$) are the nuclear spin values of the ground states of the two nuclei. 
For isotopes with even number of neutrons
and protons, we assume this value to be $I_i=0$, $I_f=0$ and for isotopes with odd number of 
neutrons and odd number of protons we take it to be $I_i=1$. $J_p$ refers to the total angular momentum of each particle in the pair, as explained 
in detail in Sec. \ref{subsec:create}. We assume that both particles in the pair have the same angular momentum and opposite projection in 
the $\hat{z}$ direction. The symmetry projection operator acts to the left on the final state, which has been studied in detail in the previous sections.

We create single-particle states with quantum numbers ($n,\ l,\  J_p,\  m_{j_p}$), where $l$ takes the values (0, 2, 4, 5)
[with $n$ respectively (0, 1, 2, 3)] \cite{Bertsch2010,Bulthuis2016}.
For instance, if $J_p=1/2$, the orbital angular momentum is $l=0$ and $n=0$. In what follows, we quote the total angular momentum value $J_p$ as shorthand.

In Eq.~(\ref{eq:transfer}) we did not include the simultaneous $(N,Z,J)$ projection since it is computationally expensive,
but we computed it for $(N,Z,J_p=1/2)$ and the qualitative trends do not change.

There are various definitions of the transfer amplitude in the literature~\cite{Grasso2012}, and given that the wave function we use is not normalized to 1,
we need to divide by the individual norms of initial and final nuclei. In addition, we are interested only in the fraction 
of the wave-function with the right ground state quantum numbers, 
so we normalize by the symmetry projected initial and final states. 

Let us turn to a detailed discussion of Fig.~\ref{fig:transfer}. 
For ${}^{132}_{66}$Dy [Fig.~\ref{fig:transfer}(a)],
the presence of spin-triplet pairing is in agreement with the addition of an $np$ pair to the lighter isotopes being highly more likely than that
of $nn$ or $pp$ pairs (which have equal transfer amplitudes). Thus, this is an additional piece of evidence on spin-triplet $np$ pairing, since ${}^{130}_{65}$Tb$\rightarrow{}^{132}_{66}$Dy is the most likely 
reaction to occur. Since ${}^{132}_{66}$Dy is an $N=Z$ nucleus, if the $np$ pairing was spin-singlet in nature, then the $np$ pair transfer would have had the same amplitude as $nn$ and $pp$ pairs.
Now turn to ${}^{132}_{64}$Gd [Fig.~\ref{fig:transfer}(b)]: the situation is rather different, since there are more neutrons 
than protons. 
Since here $nn$ and $np$ transfer amplitudes are both nonzero, we see the presence of both
spin-triplet and spin-singlet pairing. As it so happens, the spin-triplet pairing is responsible 
for the $np$ amplitude being (somewhat) larger than the $nn$ one.
Finally, in ${}^{132}_{60}$Nd [Fig.~\ref{fig:transfer}(c)] where the excess of neutrons is sizable, the situation is reversed.
We do not find any significant amplitude for $np$ or $pp$ pairs: this nucleus is characterized by spin-singlet pairing, with the most likely reaction being ${}^{130}_{60}$Nd$\rightarrow{}^{132}_{60}$Nd.



\section{Conclusions}
\label{section:Conclusions}
Symmetry restoration allows us to discern the particle number and nuclear spin eigenstate composition of the ground state
wave function found in HFB theory. By mapping out the probability 
distributions for each of these quantities for three isotopes, ${}^{132}_{66}$Dy, ${}^{132}_{64}$Gd, and ${}^{132}_{60}$Nd 
we were able to study 
how different types of spin pairings shape the eigen-composition of the ground state.
We were able to find specific patterns in the probability distributions that 
can be used as theoretical qualitative indications of spin-triplet, spin-singlet
or mixed-spin pairing. In the case of spin-triplet pairing, the proton and neutron number distributions seem rather symmetric. 
In the spin-singlet case there is checkered pattern, and the mixed-spin pairing is in between.

The second part of this work focuses on calculating ground state to ground state 
pair-transfer amplitudes in order to find the most likely candidate reactions 
for probing spin-triplet and mixed-spin pairing in
heavy nuclei. We find ${}^{130}_{65}$Tb$\rightarrow{}^{132}_{66}$Dy to be very likely, in good agreement with the spin-triplet nature of Dy.
Similarly, ${}^{130}_{60}$Nd$\rightarrow{}^{132}_{60}$Nd is the most probable transition, which is another indication of spin-singlet pairing
in this nucleus. 
The mixed-spin pairing case is more intricate, ${}^{130}_{63}$Eu$\rightarrow{}^{132}_{64}$Gd is the dominant reaction, which is an indication of spin-triplet
pairing being present, but also ${}^{130}_{64}$Gd$\rightarrow{}^{132}_{64}$Gd is likely to occur, which coincides with spin-singlet pairing. 

We are hopeful to see future experiments that can verify our predictions in this region of the nuclear chart. In addition, 
in a future work, the framework developed
and tested here, will be applied to lighter isotopes, where mixed-spin pairing might be present,
and which could be within reach of current experiments.

\begin{acknowledgments}
The authors are thankful to G. F. Bertsch and D. Lacroix 
for insightful discussions.
This work was supported
in part by the Natural Sciences and Engineering
Research Council (NSERC) of Canada, the Canada
Foundation for Innovation (CFI), the Early Researcher
Award (ERA) program of the Ontario Ministry
of Research, Innovation and Science, the
U.S. Department of Energy, Office of Science, Office of Nuclear
Physics under Contract DEAC02-05CH11231
(LBNL). Computational resources
were provided by SHARCNET and NERSC. 
\end{acknowledgments}
\appendix

%


\end{document}